%% file: art_tg_arxiv.tex
\documentclass[floatfix,aps,twocolumn,nofootinbib,superscriptaddress]{revtex4-2}

\input{header}

\usepackage{amsmath,amsfonts}
\usepackage{bm}
\usepackage{mathtools}

\usepackage{graphicx}
\usepackage{mwe}
\usepackage[version=3]{mhchem}
\usepackage{chemformula}
\usepackage{mathtools}
\usepackage{siunitx}
\usepackage{blkarray}
\usepackage{subfig}

\makeatletter
\def\maketag@@@#1{\hbox{\m@th\normalfont\normalsize#1}}
\makeatother

\DeclareMathAlphabet{\mathpzc}{OT1}{pzc}{m}{it}

\begin{document}
\captionsetup[subfigure]{labelformat=empty}
\title{Thermodynamics of growth in open chemical reaction networks}
\newcommand\unilu{\affiliation{Complex Systems and Statistical Mechanics, Department of Physics and Materials Science, University of Luxembourg, L-1511 Luxembourg City, Luxembourg}}
\newcommand\unipdchim{\affiliation{Department of Chemical Sciences, University of Padova, Via F. Marzolo, 1, I-35131 Padova, Italy}}

\author{Shesha Gopal Marehalli Srinivas}
\email{shesha.marehalli@uni.lu}
\unilu
\author{Francesco Avanzini}
\email{francesco.avanzini@unipd.it}
\unilu
\unipdchim
\author{Massimiliano Esposito}
\email{massimiliano.esposito@uni.lu}
\unilu

\begin{abstract}
We identify the thermodynamic conditions necessary to observe indefinite growth in homogeneous open chemical reaction networks (CRNs) satisfying mass action kinetics. We also characterize the thermodynamic efficiency of growth by considering the fraction of the chemical work supplied from the surroundings that is converted into CRN free energy. 
We find that indefinite growth cannot arise in CRNs chemostatted by fixing the concentration of some species at constant values, or in continuous-flow stirred tank reactors. Indefinite growth requires a constant net influx from the surroundings of at least one species. In this case, unimolecular CRNs always generate equilibrium linear growth, i.e., a continuous linear accumulation of species with equilibrium concentrations and efficiency one. Multimolecular CRNs are necessary to generate nonequilibrium growth, i.e., the continuous accumulation of species with nonequilibrium concentrations. 
Pseudo-unimolecular CRNs -- a subclass of multimolecular CRNs -- always generate asymptotic linear growth with zero efficiency. Our findings demonstrate the importance of the CRN topology and the chemostatting procedure in determining the dynamics and thermodynamics of growth.
\end{abstract}

\maketitle

%%%%%%%%%%%%%%%%%%%%%%%%%%%%%%%%%%%%%%%%%%%%%%%%%%%%%
%%%%%%%%%%%%%%%%%%%%%%%%%%%%%%%%%%%%%%%%%%%%%%%%%%%%%
%%%%%%%%%%%%%%%%%%%%%%%%%%%%%%%%%%%%%%%%%%%%%%%%%%%%%
%%%%%%%%%%%%%%%%%%%%%%%%%%%%%%%%%%%%%%%%%%%%%%%%%%%%%

Significant progress has been achieved over the last decades in establishing a thermodynamic description of open CRNs both at the stochastic and deterministic level \cite{qian2005,schmiedl2007,Polettini2014,Rao2016,Rao2018b,Avanzini2021,avanzini2022flux}. The thermodynamic framework complements the dynamical one by enabling energetic considerations of various complex chemical phenomena such as the cost of sustaining nonequilibrium steady states~\cite{gaspard2004}, chemical oscillations \cite{andrieux2008}, chaotic dynamics \cite{gaspard2020}, patterns \cite{falasco2018turing, Avanzini2020a,timur2023}, waves \cite{Avanzini2019a,kumar2021}.
It has also been used to study the efficiency of energy transduction \cite{Hill1966,wachtel2022,Corra2022}, energy storage \cite{penocchio2019eff} and various forms of information processing \cite{Amano2022,Penocchio2022}. 
But surprisingly little attention has been given to the phenomenon of chemical growth. Under which conditions does an open CRN show growth by extracting matter and energy from its surroundings and with what efficiency? The relevance of this question for biology is obvious.
Previous works on growth have mainly focused on describing the growth dynamics of CRNs with irreversible reactions \cite{Lin2020,Szathmary1991,Wills1998,Iyerbiswas2014, Nandori2022}, implicitly assuming infinite resources to grow. Thus, exponential or even hyperbolic growth was observed. In the case of reversible CRNs, numerical studies of transient growth regimes have been considered \cite{Sarkar2019}. In this Letter, we focus on the question of \textit{asymptotic} concentration growth by considering the simplest setup of homogeneous, open, and reversible CRNs in ideal dilute solutions. The  dynamics of such CRNs is given by mass action kinetics. We consider three different chemostatting mechanisms for opening CRNs: holding the concentration of some species constant (concentration control), imposing a constant influx or outflux of some species (flux control), and a constant influx and an outflux of species proportional to their concentration (mixed control). Many standard experimental setups can be described in terms of those chemostatting mechanisms, such as batch control \cite{Sorrenti2017}, continuous-flow stirred tank reactors\cite{Liu2023}, and serial transfer experiments~\cite{blokhuis2018}. By analytically studying growing unimolecular and pseudo-unimolecular CRNs and numerically studying four multimolecular CRNs, we clarify the important role of the topology of the CRN and the chemostatting procedure in determining when growth with equilibrium or nonequilibrium concentrations is possible.

%%%%%%%%%%%%%%%%%%%%%%%%%%%%%%%%%%%%%%%%%%%%%%%%%%%%%%%%%%%%%%%%
%%%%%%%%%%%%%% Main text%%%%%%%%%%%%%%%%%%%%%%%%%%%%%%%%%%%%%%%%%

We start by introducing setup and dynamics of open CRNs. Chemical species $\alpha$ are interconverted by (elementary\cite{Svehla1993}, mass balanced) chemical reactions $\rho$,
\begin{equation}
    \boldsymbol{\alpha} \cdot \boldsymbol{\nu}_{+\rho} \xleftrightharpoons[-\rho]{+\rho} \boldsymbol{\alpha}\cdot\boldsymbol{\nu}_{-\rho}\ , \
\label{Eqn:reaction_defn}
\end{equation}
where the vectors $\boldsymbol{\nu}_{\pm \rho}$ specify the number of each species participating in the reaction $\pm \rho$ and $\boldsymbol{\alpha} = (\dots,\alpha,\dots)^{\intercal}$. The topology of CRNs is encoded in the stoichiometric matrix $\mathbb{S}$, whose columns $\mathbb{S}_{\rho} \coloneqq \boldsymbol{\nu}_{-\rho} - \boldsymbol{\nu}_{+\rho}$, quantify the net change in the number of the species after the reaction $\rho$. The concentration vector, $\boldsymbol{z}(t) = (\dots,[\alpha](t),\dots)^{\intercal}$, obeys the rate equation
\begin{equation}
    \frac{d \boldsymbol{z}}{dt} = \mathbb{S}\boldsymbol{j}(\boldsymbol{z}) + \boldsymbol{I}(\boldsymbol{z}) \ . \
\label{Eqn:gen_reaction}
\end{equation}
The first term in Eq.~\eqref{Eqn:gen_reaction} is the product of the stoichiometric matrix~$\mathbb{S}$ and the reaction current vector $\boldsymbol{j}(\boldsymbol{z}) = (\dots,j_{\rho}(\boldsymbol{z}),\dots)^{\intercal}$. It tracks the total change in concentrations due to chemical reactions. Each reaction current is expressed as the difference between a forward and a backward flux,~$j_{\rho}(\boldsymbol{z}) = j_{+\rho}(\boldsymbol{z})-j_{-\rho}(\boldsymbol{z})$. From {mass action} kinetics,
\begin{equation}
    j_{\pm \rho}(\boldsymbol{z}) =  k_{\pm\rho}\boldsymbol{z}^{\nu_{\pm\rho}}\ , \
\label{Eqn:mass_action}    
\end{equation}
where $k_{\pm \rho}$ are rate constants and the notation $\boldsymbol{a}^{\boldsymbol{b}} \coloneqq \Pi_{j} a_{j}^{b_{j}}$ is used.
The second term $\boldsymbol{I}(\boldsymbol{z})$ in Eq.~\eqref{Eqn:gen_reaction} results from chemostatting a subset of species (denoted $\textit{Y}$) and quantifies the exchange currents with the surroundings. The remaining species (denoted $\textit{X}$) are called internal species. As a result, Eq.~\eqref{Eqn:gen_reaction} splits into 
$d_{t}\boldsymbol{x} = \mathbb{S}^{X}\boldsymbol{j}(\boldsymbol{z})$ and $d_{t}\boldsymbol{y} = \mathbb{S}^{Y}\boldsymbol{j}(\boldsymbol{z}) + \boldsymbol{I}^{Y}(\boldsymbol{z})$, where $\boldsymbol{x}$ and $\boldsymbol{y}$ represent the concentrations of the internal and chemostatted species, respectively.  
The three types of chemostatting procedures we consider are formalized as follows. 
\textit{Concentration control} maintains $\boldsymbol{y}$ constant in time: $\boldsymbol{I}^{Y}(\boldsymbol{z}) = -\mathbb{S}^{Y}\boldsymbol{j}(\boldsymbol{z})$. \textit{Flux control} imposes a constant flux of species into or out of the system: $\boldsymbol{I}^{Y}(z) = \Tilde{\boldsymbol{I}}$. \textit{Mixed control} imposes a constant influx of species into the system and extracts species from the system proportionally to their concentrations: $\boldsymbol{I}^{Y}(z) = -\Tilde{\mathbb{D}}\boldsymbol{y}+\Tilde{\boldsymbol{I}}$, where $\Tilde{\mathbb{D}}$ is a diagonal matrix and $\Tilde{\boldsymbol{I}}$ is a constant (positive) vector. Continuous-flow stirred tank reactors correspond to  a form of mixed control in which all the species are extracted with the same rate, i.e., $\Tilde{\mathbb{D}} = {k}^{e}\mathbb{1}$ \cite{blokhuis2018}.
From Eqs.~\eqref{Eqn:gen_reaction} and~\eqref{Eqn:mass_action}, we note that unimolecular CRNs, which consist exclusively of reactions of the form $\alpha \xrightleftharpoons{} \beta$, will follow a linear dynamics for any chemostatting procedure. CRNs with multimolecular reactions will always give rise to nonlinear dynamics for flux and mixed control. However, under concentration control, some of them may give rise to linear dynamics. We call these pseudo-unimolecular CRNs.

We now turn to the thermodynamics of open CRNs \cite{Rao2016,Avanzini2021,avanzini2022flux}. The chemical potential for species in homogeneous dilute solutions at temperature $T$ is given by the vector $\boldsymbol{\mu}(\boldsymbol{z}) = \boldsymbol{\mu}^{0} + RT\text{ln}(\boldsymbol{z})$, with $R$ the gas constant and $\boldsymbol{\mu}^{0}$ the vector of standard chemical potentials. From the form of the chemical potential, we deduce the nonequilibrium Gibbs free energy,~$G(\boldsymbol{z}) = \sum_{\alpha} \mu_{\alpha}[\alpha] - RT\sum_{\alpha}[\alpha]$, and the entropy production rate (EPR),~$T\dot{\Sigma} = -\boldsymbol{\mu}\mathbb{S}\boldsymbol{j}$. The thermodynamics is linked to the dynamics via the \textit{local detailed balance},
\begin{equation}
     RT~\text{ln}\left({k_{+\rho}}/{k_{-\rho}}\right) = -\boldsymbol{\mu}^{0} \cdot \mathbb{S}_{\rho} \ . \
\label{Eqn:LDB}
\end{equation}
This allows us to express the EPR as  $\dot{\Sigma} =   {R}\sum_{\rho} (j_{+\rho}-j_{-\rho} ) \text{ln} (j_{+\rho}/j_{-\rho}) \geq 0$, and derive a nonequilibrium second law~$T\dot{\Sigma} = \dot{w}_{\text{c}}-d_{t}{G} \geq 0 $. The term~$\dot{w}_{\text{c}} = \sum_{\alpha} \mu_{\alpha}I_{\alpha}$, called the chemical work rate, is the total work done on the CRN per unit time. Thus, the second law states that the dissipation is the difference between the total work done on the CRN and the change in its free energy.   
We further decompose the chemical work rate~$\dot{w}_{c}$ using \textit{conservation laws}, which are linearly independent left null eigenvectors $\ell^{\lambda}$ of the stoichiometric matrix~$\mathbb{S}$. They identify moieties, that are parts of (or entire) molecules that are left intact by the reactions. Their concentrations ${L}^{\lambda} = \ell^{\lambda}\cdot\boldsymbol{z}$ are conserved in closed CRNs. In open CRNs,
\begin{equation}
     d_{t}{L}^{\lambda} = \ell^{\lambda}\cdot d_{t}\boldsymbol{z} = \ell^{\lambda}\cdot\boldsymbol{I}(z)\\.
\label{Eqn:Mass_Evol}
\end{equation}
If $\ell^{\lambda} \cdot \boldsymbol{I}(z) \neq 0$ (resp. $= 0$), the moiety concentration ${L}^{\lambda}$ is no longer conserved (resp. still conserved) and the conservation law $\ell^{\lambda}$ is said to be \textit{broken} (resp. \textit{unbroken}).
Note that every CRN has at least one conservation law that involves all species, denoted $\ell^{m}$, with the corresponding moiety concentration~$L^{m} = \ell^{m} \cdot\boldsymbol{z}$ called the mass density~\cite{Angeli2009}, which is always broken in an open CRN.

Since the moiety concentrations ${L}^{\lambda}$ change only due to the exchange currents, we split the chemical work rate into the work done in changing ${L}^{\lambda}$ and the remaining. However, as chemostatting a species does not always break a conservation law,  we first split the set of chemostatted species~$Y$ into potential species~$Y_{p}$ which break conservation laws and force species~$Y_{f} = Y\setminus Y_{p}$. 
This allows us to associate a moiety to a single $Y_p$ species and express its concentration according to 
$\boldsymbol{m}(\boldsymbol{z}) = \big(\mathbb{L}^{b}_{Y_{p}}\big)^{-1}\mathbb{L}^{b}\cdot\boldsymbol{z}$, where $\mathbb{L}^{b}$ is a matrix with broken conservation laws as rows while $\mathbb{L}^{b}_{Y_{p}}$ is its square submatrix~\cite{Avanzini2021}. 
The work done in changing the exchanged moiety concentrations, called the moiety work rate, then reads~$\dot{w}_{\text{m}} = \boldsymbol{\mu}_{Y_{p}}d_{t}\boldsymbol{m}$. The remaining part of the chemical work rate is called the nonconservative work rate, $\dot{w}_{\text{nc}} = \dot{w}_{\text{c}}- \dot{w}_{\text{m}}$. The nonconservative work rate identifies the energetic cost of maintaining fluxes
of the same moiety between the various chemostats \cite{Rao2016, Avanzini2021} and can be rewritten as~$\dot{w}_{\text{nc}}= \big(\boldsymbol{\mu}_{Y}-\boldsymbol{\mu}_{Y_{p}}\big(\mathbb{L}^{b}_{Y_{p}}\big)^{-1}\mathbb{L}^{b}\big)\cdot\boldsymbol{I}^{Y}$. As a result, the nonequilibrium second law of CRNs can be rewritten in a form particularly well suited to analyzing growth:
\begin{equation}
    T\dot{\Sigma} = \dot{w}_{\text{nc}}+\dot{w}_{\text{m}} -d_{t}{G} \geq 0 \;.
\label{Eqn:Second_Law}    
\end{equation}

Closed CRNs always reach a fixed point, $\boldsymbol{z}_{\text{eq}}$, called the \textit{equilibrium} state, such that $\boldsymbol{j}(\boldsymbol{z}_{\text{eq}}) = 0$ and $T\dot{\Sigma} = \dot{w}_{\text{nc}} = 0$. Fixed points in open CRNs can either be equilibria or \textit{nonequilibrium steady states} (NESS),~ $\boldsymbol{z}_{\text{ss}}$, with $\boldsymbol{j}(\boldsymbol{z}_{\text{ss}}) \neq 0$ and $T\dot{\Sigma} = \dot{w}_{\text{nc}} > 0$. Note, for any fixed point of a CRN, $\dot{w}_{\text{m}} = d_{t}G = 0$. Often, chemostatting CRNs is done by concentration control. Generically, such open CRNs relax to a  NESS or exhibit complex behaviors like limit cycles or multistability. However, no asymptotic growth was observed till recently, in Ref. {\cite{avanzini2022flux}}, where numerically,  under both flux and mixed control, a CRN was shown to tend towards a state with infinite concentrations. In this work, we investigate the dynamics and energetics of growing CRNs.

Indefinite growth implies that at least one species keeps accumulating in the system. It is thus equivalent to the statement that  $\lim_{t \to \infty} ||\boldsymbol{z}(t)-\boldsymbol{z}(0)|| = \infty$ (where $||\bullet||$ denotes the Euclidean norm), or that the mass density diverges $\lim_{t \to \infty} L^{m}(t) = \infty$. Thermodynamically, the accumulation of species results in free energy that diverges in time. This tells us that the work needed from the surroundings also has to diverge which immediately implies that closed CRNs do not grow and only open CRNs may grow. The chemostatting procedure plays a key role in determining the growth conditions. For example, in the case of continuous-flow stirred tank reactors, Eq.~\eqref{Eqn:Mass_Evol} for the mass density becomes $d_{t}{L}^{m} = -{k}^{e}{L}^{m} + \ell^{m}_{Y}\cdot\Tilde{\boldsymbol{I}}$, implying a bounded solution and therefore no growth for any choice of rate constants~$\{k_{\pm \rho},{k}^{e}\}$ and current $\Tilde{\boldsymbol{I}}$~\cite{blokhuis2018}. As another example, under flux control, all CRNs show growth if $d_{t}{L}^{m} = \ell^{m}\cdot\Tilde{\boldsymbol{I}} > 0$, with all growing concentrations diverging at most linearly in time. 

For a systematic analysis, we first consider CRNs that are dynamically linear and can thus be characterized analytically. 
We note that the detailed calculations and a more extensive analysis are reported in a companion paper \cite{Comp}.
The first important result is that unimolecular CRNs can produce equilibrium growth under flux control and no growth otherwise. 
For any chemostatting procedure of a unimolecular CRN, Eq.~\eqref{Eqn:gen_reaction} can be rewritten in the form
\begin{equation}
    d_{t}\boldsymbol{a} = \mathbb{W}\boldsymbol{a} -\mathbb{D}\boldsymbol{a} + \bar{\boldsymbol{I}} \,.
\label{Eq:linear_general}     
\end{equation}
Here,  $\boldsymbol{a} = \boldsymbol{z}$ under flux and mixed control, while $\boldsymbol{a} = \boldsymbol{x}$ under concentration control.
$\mathbb{W}$ is a rate matrix \cite{vankampen}  
whose nondiagonal entries 
are given by $\mathbb{W}_{\alpha, \beta} = \{\sum_{\rho} k_{+\rho} ~~ \big{|}~\rho: \beta \xrightleftharpoons{+\rho}\alpha \}$ 
and diagonal entries are such that the sum over each column is zero.
Due to the local detailed balance condition~\eqref{Eqn:LDB}, $\mathbb{W}$ is a detailed balanced matrix with a zero eigenvector proportional to $\exp(-{\boldsymbol{\mu^{0}}}/{{RT}})$. 
% -
Under all kinds of control, $\bar{\boldsymbol{I}}$ is a constant vector.
% -
Under mixed control or concentration control, $\mathbb{D}$ is a non-negative diagonal matrix with at least one nonzero element. 
As a result, all the eigenvalues of the matrix $\mathbb{W}-\mathbb{D}$ are negative and growth cannot occur.

Flux control instead corresponds to $\mathbb{D} = 0$ in Eq.~\eqref{Eq:linear_general}. 
The mass density grows linearly in time at a rate given by the total injection current, $d_t L^{m}(t) = \sum_{\alpha}\bar{{I}}_{\alpha}$, and the growth of all species is controlled by it at large times:
\begin{equation}
    [\alpha](t) = \underbrace{\left(\frac{e^{{-\mu^{0}_{\alpha}}/{{RT}}}}{\sum_{\alpha'}e^{{-\mu^{0}_{\alpha'}}/{{RT}}}}\right) L^{m}(t)}_{= [\alpha]_{\text{eq}}\left(L^{m}(t)\right)} + ~ c_{\alpha}(\bar{\boldsymbol{I}}) = \mathcal{O}(t)\ . \  \label{Eqn:Linear_conc_growth}
\end{equation}
Here, $\{c_{\alpha}(\bar{\boldsymbol{I}})\}$ are time-independent coefficients
and linear functions of the injection currents, while $[\alpha]_{\text{eq}}\left(L^{m}(t)\right)$ is the equilibrium value to which the CRN would relax if it were closed (by stopping the injection) at time $t$.   
This means that the growth process follows, up to the offset $c_{\alpha}(\bar{\boldsymbol{I}})$, that equilibrium value.
Turning to the second law, at large times one finds that 
\begin{align}
    T\dot{\Sigma} &\sim \dot{w}_{\text{nc}} = \mathcal{O}\left(t^{-1}\right), \label{Eqn:Thermo_diss_linear_growth}\\
    d_{t}{G} &\sim \dot{w}_{\text{m}} = \mathcal{O}\left(\text{ln}\left(t\right)\right) . \label{Eqn:Thermo_Gibbs_linear_growth}
\end{align}
The physical interpretation of this result is simple. Since the CRN concentrations scale extensively over time,  while the offset in Eq.~\eqref{Eqn:Linear_conc_growth} remains intensive, growth over large times resembles a quasistatic process where the total dissipation becomes negligible and the moeity work performed by the injection is reversibly converted into free energy. Defining the thermodynamic efficiency with which the chemical work is converted into free energy, $\eta = d_{t}G/\dot{w}_{\text{c}}$, and using Eqs.~\eqref{Eqn:Thermo_Gibbs_linear_growth},~\eqref{Eqn:Thermo_diss_linear_growth} and \eqref{Eqn:Second_Law}, we see that $\eta$ tends to one at large times for unimolecular CRNs, indicating reversible energy conversions. In other words, equilibrium growth can be seen at large times as a moving equilibrium following the mass density displaced by the injection. 

We now analyze pseudo-unimolecular CRNs. 
These are multimolecular CRNs where the concentrations of certain species, called hidden species, are held constant via concentration control, and where the dynamics of the remaining concentrations, called dynamically linear species, become linear. For example, consider the single reaction $\text{E} + \text{S} \xrightleftharpoons[]{} \text{ES}$. If the concentration $[\text{S}]$ is held constant, the dynamics becomes linear. By absorbing the (constant) concentrations of the hidden species into the rate constants, the rate equation for pseudo-unimolecular CRNs can still be written as Eq.~\eqref{Eq:linear_general} with a rate matrix $\mathbb{W}$ that is not detailed balanced. 
For the same reasons as for unimolecular CRNs, mixed control and concentration control on the dynamically linear species will not give rise to any growth. Instead, flux control on the dynamically linear species will produce growth which over long times reads
\begin{equation}
    [\alpha](t) =\underbrace{\pi^{\text{ss}}_{\alpha} L^{m}(t)}_{=[\alpha]_{\text{ss}}\left(L^{m}(t)\right)} + ~c_{\alpha} .
\label{Eqn:PseudoLinear_conc_growth}
\end{equation}
Here, $\pi^{\text{ss}}_{\alpha}$ is a component of the normalized right null vector of $\mathbb{W}$ and $[\alpha]_{\text{ss}}\left(L^{m}(t)\right)$ is the steady state (generically a NESS) to which the CRN would relax if it were closed (by stopping the injection) at time $t$.
In the long time limit, the thermodynamic quantities are of the form,
\begin{align}
     T\dot{\Sigma} &\sim \dot{w}_{\text{nc}} = \mathcal{O}(t)\,,\label{Eq:Thermo_diss_pseudo_linear}\\
     d_{t}{G} &\sim \dot{w}_{\text{m}} = \mathcal{O}\left(\text{ln}\left(t\right)\right)\;.
    \label{Eq:Thermo_driving_pseudo_linear}
\end{align}
Exactly as for growing unimolecular CRNs (see  Eq.~\eqref{Eqn:Thermo_Gibbs_linear_growth}) the moiety work rate performed by the injection is converted into free energy. 
The key difference is that this conversion is generically irreversible, and thus gives rise to an extensive EPR in time. In contrast to the unimolecular case, the efficiency $\eta = d_{t}G/\dot{w}_{\text{c}}$ tends to zero at large times (using Eqs.~\eqref{Eq:Thermo_diss_pseudo_linear}, \eqref{Eq:Thermo_driving_pseudo_linear} and \eqref{Eqn:Second_Law}) indicative of very inefficient irreversible conversions. This non-equilibrium growth process can be seen over long times as a moving NESS following the mass density displaced by the injection. The dissipation is essentially the nonconservative work dissipated in the NESS due to the presence of the chemostatted hidden species.

\begin{figure*}
\begin{minipage}[b]{.4\textwidth}
\centering
\includegraphics[width = \columnwidth]{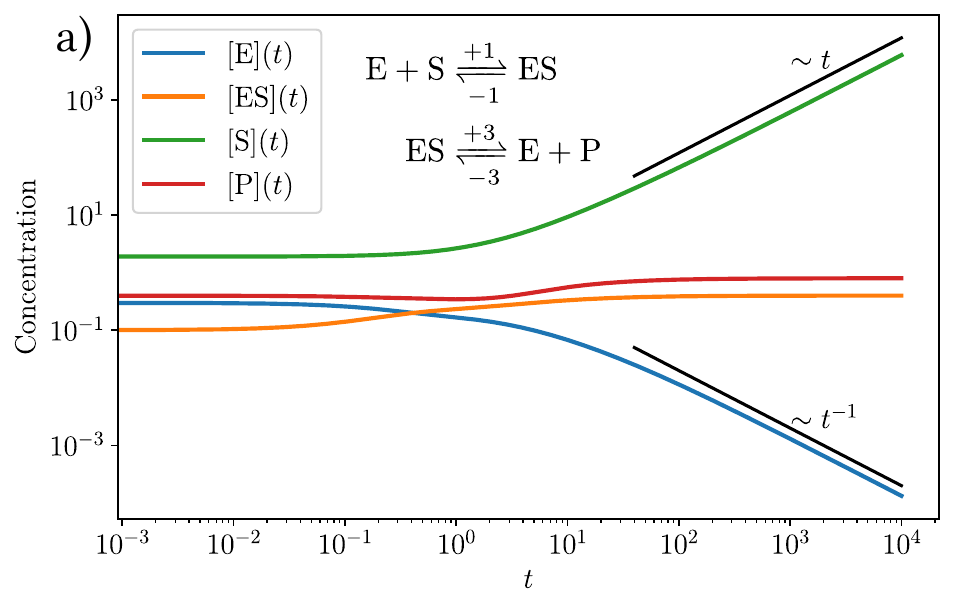}
\end{minipage}\qquad
\begin{minipage}[b]{.4\textwidth}
\centering
\includegraphics[width = 0.97\columnwidth]{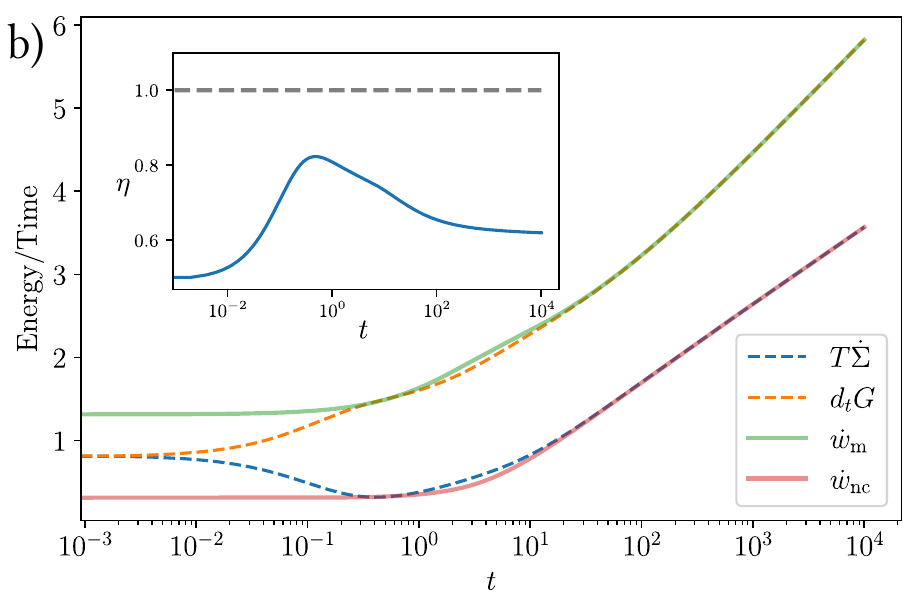}
\end{minipage}
\caption{Evolution of a) the concentrations, b) the contributions to the EPR in Eq.~\eqref{Eqn:Second_Law} and the efficiency $\eta$, for the Michaelis Menten CRN in a) under mixed control when $\text{S}$ is injected at the rate $I_{\ch S} =1$ while $\text{P}$ is extracted at the rate $-{k}^{e}_{\ch{P}}[\ch{P}]$. We rescaled time, concentrations, and energy/time by $1/k_{-1}$, $k_{-1}/k_{+1}$ and $RT k_{-1}^{2}/k_{+1}$, respectively. $[\text{E}](0)=0.3$, $[\text{ES}](0)=0.1$, $[\text{S}](0)=1.9$, $[\text{P}](0)=0.4$, ${k}^{e}_{\ch{P}} = 0.5$, $k_{\pm 1} = k_{\pm 2} = 1$, $\mu^{0}_{\ch{E}} = 1$, $\mu^{0}_{\ch{ES}} = 2$, $\mu^{0}_{\ch{S}} = 1$ and $\mu^{0}_{\ch{P}} = 1$. }
\label{fig:Michealis_Menten}
\end{figure*}

In summary, by considering dynamically linear CRNs, we prove that only flux chemostatting generates growth in such CRNs. Unimolecular CRNs only display equilibrium growth, while pseudo-unimolecular CRNs can give rise to nonequilibrium growth.
A striking feature of these CRNs is a splitting of the second law, where the total dissipation is essentially the nonconservative work rate whereas the moiety work rate is essentially converted into free energy.

We now turn to multimolecular CRNs. General statements are notoriously difficult to make in this case. However, it is conjectured that multimolecular CRNs cannot grow under concentration control. Although this is proved for the cases of single linkage class CRNs~\cite{anderson2011} and strongly endotactic CRNs~\cite{manoj2014}, a general proof is still lacking. 
We considered four different multimolecular CRNs, based on their different topological properties, which are analyzed (analytically and numerically) in detail in the companion paper \cite{Comp}. 
Two of them are shown in the present letter. All support the conjecture. The intuition behind this conjecture is that the accumulation of any species will eventually produce reaction currents (due to mass action kinetics) which will consume the species and these currents would be sufficiently large to balance the external currents generated by the chemostats.
Using these four models, we also found that, contrary to dynamically linear CRNs, some multimolecular CRNS can grow under mixed control.
Indeed, in Fig.~\ref{fig:Michealis_Menten}a, we show that the Michaelis Menten CRN under mixed control grows out of equilibrium.

\begin{figure*}
\begin{minipage}[b]{.4\textwidth}
\centering
\includegraphics[width = \columnwidth]{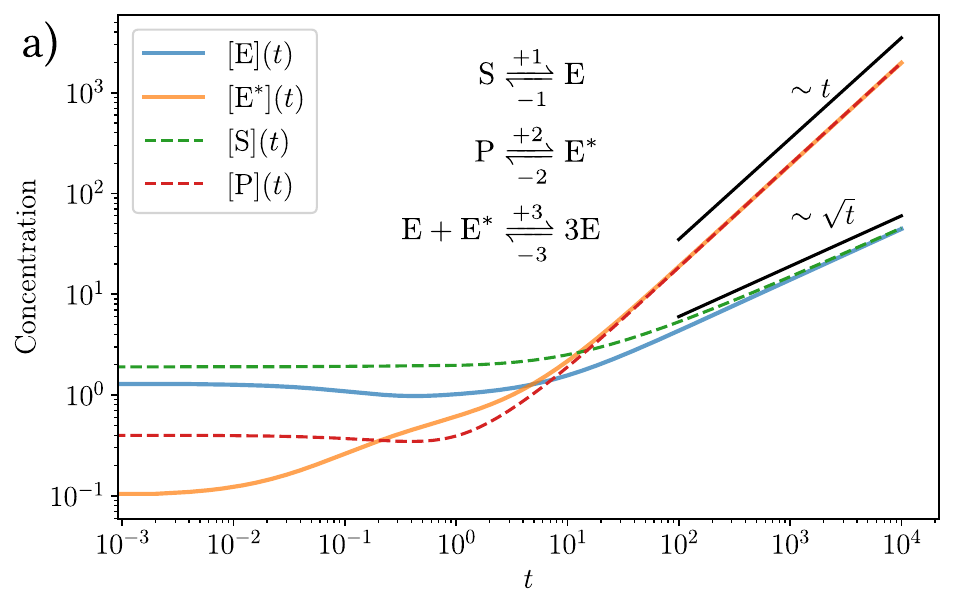}
\end{minipage}\qquad
\begin{minipage}[b]{.4\textwidth}
\centering
\includegraphics[width = \columnwidth]{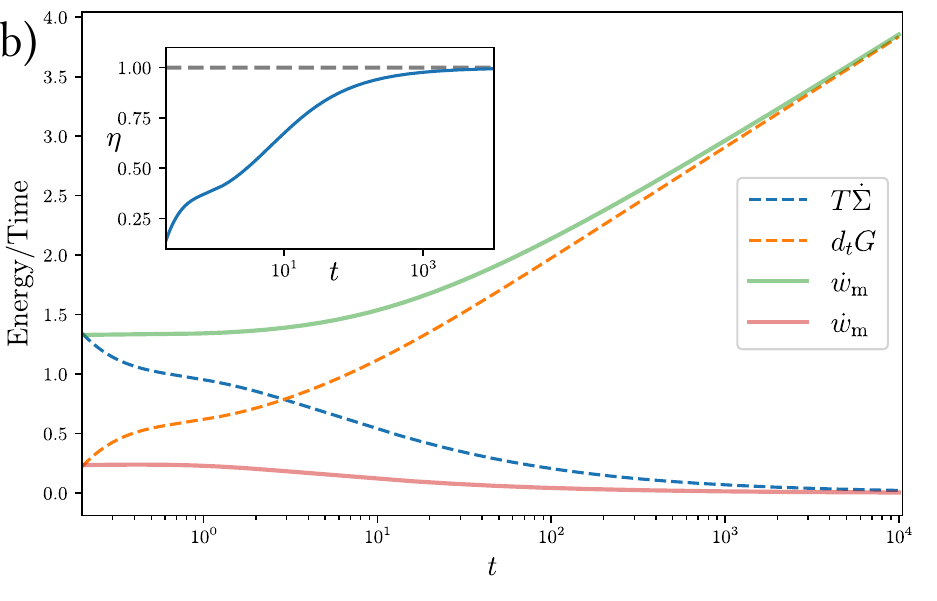}
\end{minipage}
\caption{Evolution of a) the concentrations, b) the contributions to the EPR in Eq.~\eqref{Eqn:Second_Law} and the efficiency $\eta$, for the autocatalytic CRN in a) under flux control when $\text{S}$ is injected at the rate $I_{\ch S} = 1$ while $\text{P}$ is extracted at the rate $I_{\ch P} = -0.1$. Here, we rescale time, concentrations, and energy/time by $1/k_{+1}$, $k_{+1}/k_{+3}$ and $RT k_{+1}^{2}/k_{+3}$, respectively. $[\ch{E}](0) = 1.3$, $[\ch{E}^{*}](0)=0.1$, $[\ch{F}](0)=1.9$, $[\ch{W}](0)=0.4$, $k_{\pm 1}=k_{\pm 2} = k_{\pm 3} =1$, $\mu^{0}_{\ch{E}} = 1$, $\mu^{0}_{\ch{E^{*}}} = 2$, $\mu^{0}_{\ch{S}} = 1$, $\mu^{0}_{\ch{P}} = 2$.}
\label{fig:Autocat_core}
\end{figure*}

For all four models, irrespective of the type of control, in growth regimes, we observed that there exists a choice of potential species such that the thermodynamics displays the same splitting of the second law seen in unimolecular and pseudo-unimolecular CRNs (see Eqs.~\eqref{Eqn:Thermo_diss_linear_growth},~\eqref{Eqn:Thermo_Gibbs_linear_growth},  ~\eqref{Eq:Thermo_diss_pseudo_linear} and ~\eqref{Eq:Thermo_driving_pseudo_linear}) of the form 
\begin{align}\label{Eq:Thermo_diss_growth}
       T\dot{\Sigma} &\sim \dot{w}_{\text{nc}} \geq 0\,,\\
       d_{t}G        &\sim \dot{w}_{\text{m}} > 0 \label{Eq:Thermo_driving_growth}\,.
\end{align}
However, the generality of this observation remains to be tested beyond these models. 
Furthermore, the energy conversions in growing multimolecular CRNs are generically such that the efficiency might be strictly between zero and one (see inset of Fig.~\ref{fig:Michealis_Menten}b).
Finally, multimolecular CRNs can display growth regimes that are significantly different from those in dynamically linear CRNs. Only a subset of species may grow and the temporal scalings of the concentrations may differ, as shown in Fig.~\ref{fig:Autocat_core}a.

\begin{figure}[!t]
    \centering   
    \includegraphics[width = \columnwidth]{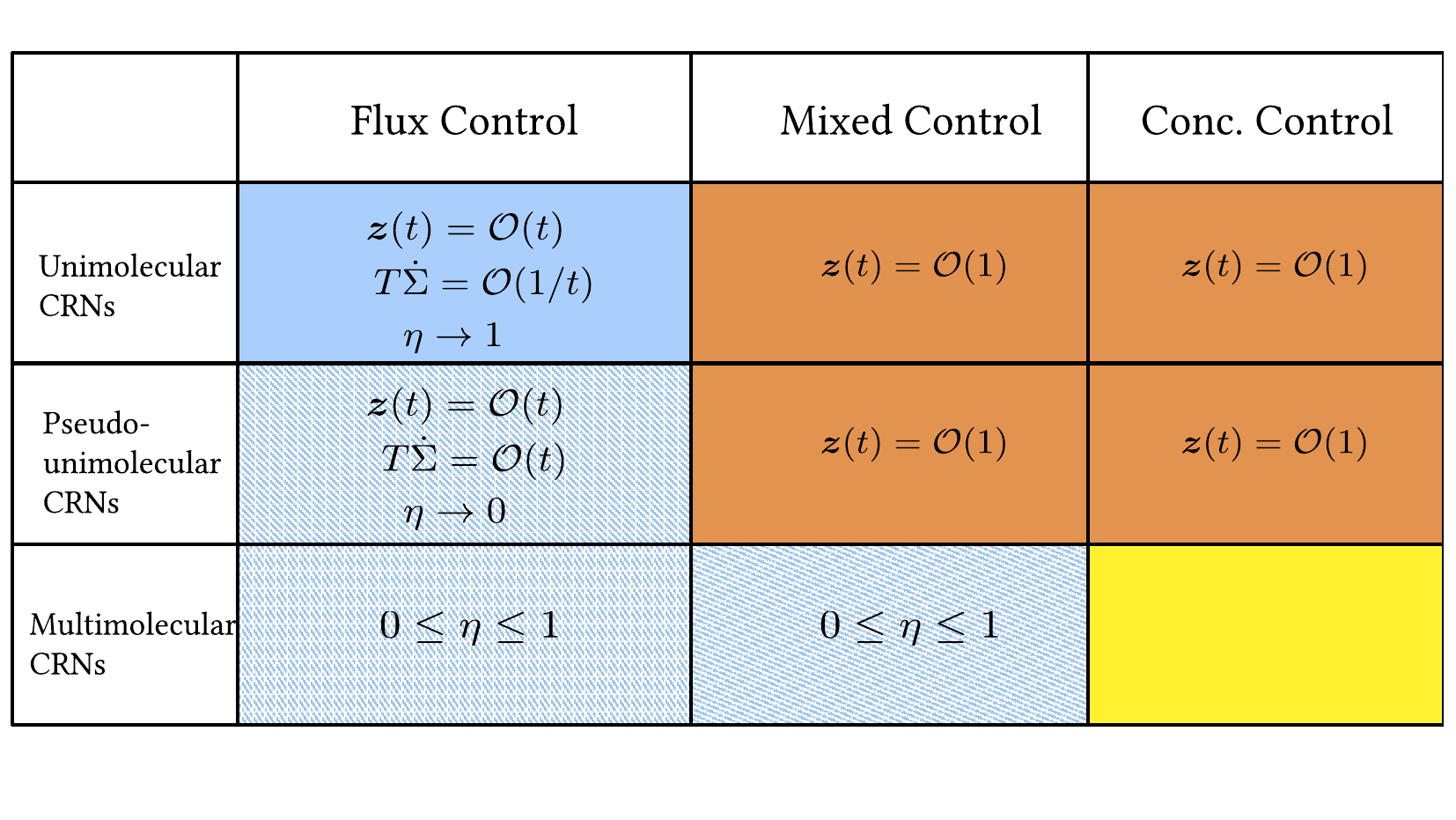}
    \caption{Long-time behavior of concentrations, EPR, and efficiency. 
    No growth under orange (resp. yellow) conditions (resp., but definite proof is still missing).
    Equilibrium (resp. nonequilibrium) growth under solid (resp. striped) blue conditions.
    }
    \label{fig:Results_tabular}
\end{figure}
{\it Conclusion and perspectives} -
Our main findings are summarized in Fig.~\ref{fig:Results_tabular}. We considered growth in dilute ideal solutions, implicitly assuming that over the timescales considered, the species concentrations remain negligible compared to the solvent.
Approaches to growth introducing an equation of state coupling volume and species concentrations have been considered but only for detailed balanced dynamics (which always relax to equilibrium) \cite{Kobayashi2022,bigan2015,Kondo2011}. A future perspective may be to consider the energetics of such systems for non-detailed balanced dynamics.
Characterizing the effect of diffusion or the effect of compartment division \cite{bauermann2022,klosin2020,Ianeselli2022} on growth may also be considered. 
Finally, we observed that autocatalysis does not seem to play an important role on the long time growth (see companion paper \cite{Comp}). Perhaps the role of autocatalysis is more apparent in the short-time growing behavior~\cite{unterberger2021, blokhuis2020, Sarkar2019}.

\textit{Acknowledgements} - 
The authors would like to thank Danilo Forastiere and Gianmaria Falasco for helpful discussions. 
This research was supported by the Luxembourg National
Research Fund (FNR), via the research funding schemes PRIDE
(Grant No. 19/14063202/ACTIVE), CORE project ChemComplex (Grant No. C21/MS/16356329), and by project INTER/FNRS/20/15074473 funded by F.R.S.-FNRS (Belgium) and FNR (Luxembourg).

\bibliography{biblio}

\end{document}

%% file: header.tex
\usepackage[utf8]{inputenc}
\usepackage[lining,semibold]{libertine} % a bit lighter than Times--no osf in math
\usepackage[libertine, cmintegrals, bigdelims, vvarbb]{newtxmath}
\usepackage{amsmath}
\usepackage{amsfonts}
\usepackage{mathrsfs}
\usepackage{gensymb}
\usepackage{bbm}
\usepackage{dsfont}

\usepackage{kbordermatrix}
\renewcommand{\kbldelim}{(}
\renewcommand{\kbrdelim}{)}

\usepackage{chemformula}
\usepackage[caption=false]{subfig}
\usepackage{chemfig}
\usepackage[version=3]{mhchem}

\usepackage{tikz}
\usetikzlibrary{matrix,positioning,decorations.pathreplacing}

\usepackage{soul}
\usepackage{xcolor}

\usepackage{scalerel} % To change size things in math environment \scaleto{...}{1pt}

\definecolor{webgreen}{rgb}{0,.5,0}
\definecolor{webbrown}{rgb}{.6,0,0}
\definecolor{grigio}{rgb}{.85,.85,.85} 
\definecolor{RoyalBlue}{rgb}{0.0, 0.14, 0.4}
\definecolor{skyblue1}{rgb}{0.45,0.62,0.81}
\definecolor{skyblue2}{rgb}{0.2,0.39,0.64}
\definecolor{skyblue3}{rgb}{0.13,0.29,0.53}
\definecolor{scarlet1}{rgb}{0.93,0.16,0.16}
\definecolor{scarlet2}{rgb}{0.8,0,0}
\definecolor{scarlet3}{rgb}{0.64,0,0}

\definecolor{g}{gray}{0.50}

\usepackage{hyperref}
\hypersetup{%
    colorlinks=true, linktocpage=true, pdfstartpage=1, pdfstartview=FitV,%
    breaklinks=true, pdfpagemode=UseNone, pageanchor=true, pdfpagemode=UseOutlines,%
    plainpages=false, bookmarksnumbered, bookmarksopen=true, bookmarksopenlevel=1,%
    hypertexnames=true, pdfhighlight=/O,%nesting=true,%frenchlinks,%
    urlcolor=webbrown, linkcolor=RoyalBlue, citecolor=webgreen, %pagecolor=RoyalBlue,%
    pdftitle={},%
    pdfauthor={Francesco Avanzini},%
    pdfsubject={},%
    pdfkeywords={},%
    pdfcreator={pdfLaTeX},%
    pdfproducer={LaTeX REVTeX}%
}